# Machine and deep learning methods for predicting 3D genome organization

Brydon P. G. Wall[1], My Nguyen[2], J. Chuck Harrell[3,4,5], Mikhail G. Dozmorov[2,3*]

[1]Center for Biological Data Science, Virginia Commonwealth University, Richmond, VA, 23284, USA.

[2]Department of Biostatistics, Virginia Commonwealth University, Richmond, VA, 23298, USA.

[3]Department of Pathology, Virginia Commonwealth University, Richmond, VA, 23284, USA.

[4]Massey Comprehensive Cancer Center, Virginia Commonwealth University, Richmond, VA 23298, USA.

[5]Center for Pharmaceutical Engineering, Virginia Commonwealth University, Richmond, VA 23298, USA.

[*]To whom correspondence should be addressed: mdozmorov@vcu.edu

**Abstract**

Three-Dimensional (3D) chromatin interactions, such as enhancer-promoter interactions (EPIs), loops, Topologically Associating Domains (TADs), and A/B compartments play critical roles in a wide range of cellular processes by regulating gene expression. Recent development of chromatin conformation capture technologies has enabled genome-wide profiling of various 3D structures, even with single cells. However, current catalogs of 3D structures remain incomplete and unreliable due to differences in technology, tools, and low data resolution. Machine learning methods have emerged as an alternative to obtain missing 3D interactions and/or improve resolution. Such methods frequently use genome annotation data (ChIP-seq, DNAse-seq, etc.), DNA sequencing information (k-mers, Transcription Factor Binding Site (TFBS) motifs), and other genomic properties to learn the associations between genomic features and chromatin interactions. In this review, we discuss computational tools for predicting three types of 3D interactions (EPIs, chromatin interactions, TAD boundaries) and analyze their pros and cons. We also point out obstacles of computational prediction of 3D interactions and suggest future research directions.

**Keywords:** Hi-C, enhancer-promoter interactions, chromatin, loops, TADs, machine learning, deep learning, software

**Introduction**

Historically, biological characterization of a genome began with the investigation of individual genomic elements, such as genes, Transcription Factor Binding Sites (TFBSs), histone modifications, CpG sites, and DNA methylation [1], collectively referred to as the 1D genome organization. The advent of chromosome conformation capture sequencing technologies [2–4] revealed a distinct, complementary aspect of genome organization - its Three-Dimensional (3D) structure. These technologies have revealed important 3D spatial constructs such as chromosomal territories, A/B compartments [5], topologically associating domains (TADs) [6,7], and at the most local scale, chromatin loops and enhancer-promoter interactions (EPIs) [8–10].

TADs are self-interacting structural domains within a genome formed through loop extrusion. The cohesin complex is first loaded onto DNA, then extrudes the DNA until it reaches a barrier, usually with the convergent orientation of CTCF and several other DNA-binding proteins [11,12,8,13,14]. Loops and EPIs frequently occur within TADs and are manifested as "point-to-point" interactions facilitating regulatory processes. Emerging evidence has linked chromatin interactions to critical roles in cell dynamics, such as gene regulation and cell differentiation. An example of this is the mechanism by which distal enhancers come into contact with promoters to initiate transcription [15–17]. Consequently, the disruption of the higher-order chromatin structures has been shown to lead to developmental diseases [18], cancer [19–22] and other disorders [23,24]. Although subsets of 3D structures are tissue-invariant [6,8,7], many are cell-type-specific [5,25] representing potential targets for therapeutic intervention. Therefore, defining precise locations of 3D chromatin structures is a top priority in understanding the biology of genome regulation in health and disease.

Hi-C, a high-throughput chromatin conformation capture technology, allows for the detection of chromatin interactions on a genome-wide scale. However, Hi-C technology has its limitations in the analysis of 3D chromatin structures. A typical Hi-C experiment requires billions of reads, more than 20 times the amount of sequencing of a typical RNA-seq experiment [8]. This leads to a prohibitively high cost for this use case. Also, cross-linked DNA fragments are digested with restriction enzymes cutting at predefined sites. Enzymes recognizing 6 bp sequences cut the DNA into fragments with an average size of about 4 Kb, which is suboptimal. Even with 4-cutter restriction enzymes theoretically cutting at every 256 bp, the uneven distribution of cutting sites across a genome makes the resolution variable between individual fragments [26]. A more common approach is to bin a genome into regions of equal size (typically, 5Kb - 100Kb) and count the number of interactions between regions. Because increasing the resolution of Hi-C data requires a quadratic increase in the total sequencing depth [27], high-resolution Hi-C data (1 Kb [28,8], or even 750 bp [29]) remains rare. Furthermore, the lack of systematic collections of high-resolution Hi-C data is compounded by a wide variety of chromatin conformation capture protocols. This further hinders the detection and comparison of higher-order chromatin structures.

Numerous studies have demonstrated the association of higher-order chromatin structures with genomic annotations and sequence features [25,30,31,9,32], reviewed in [33–36]. This collection of genome annotation data, which are regions annotated as carrying functional/regulatory potential or having a biological property, are collectively referred to hereafter as epigenomic data [37]. A/B compartments were found to correspond to transcriptionally active or inactive chromatin regions, enrichment or depletion in DNA accessibility signal, gene density, and replication timing [5,38]. Boundaries of TADs were found to be enriched in CTCF (considering the convergent orientation of CTCF binding motifs) and other architectural proteins of cohesin and mediator complex (RAD21, SMC3, and others) [39,40,14,41], marks of transcriptionally active chromatin (DNAse hypersensitive sites, H3K4me3, H3K27ac, and H3K36me3 histone modifications) [42–44,7], and housekeeping genes [45,6]. Similarly, smaller chromatin loops were found to be mediated by CTCF and cohesin via a loop extrusion mechanism, where CTCF binds to a specific and non-palindromic motif in a convergent orientation at two sites acting as loop anchors [11,46,6,14,47]. These relationships suggest that genomic annotations and sequence features may be used to predict the location of higher-order chromatin structures.

In contrast to limited 3D data, the amount of 1D genome annotation data, also known as epigenomic data, has been growing exponentially. ENCODE, NIH Roadmap Epigenomics, FANTOM5, BLUEPRINT, and other members of the International Human Epigenome Consortium (IHEC) [48] have been actively cataloging functional and regulatory genome annotation datasets. These datasets include cell-type and tissue-specific histone modifications, DNAse I hypersensitive sites, DNA methylation, and TFBSs. These data are typically available at a resolution much higher than 3D data (tens of bases in contrast to tens of kilobases) which provides opportunities to refine the location of higher-order chromatin structures.

This review focuses on the machine learning methods and tools for predicting higher-order chromatin structures using genomic annotations, DNA sequences, and other genomic properties (Figure 1). We first describe the machine learning framework and considerations needed for predicting chromatin interactions. We then describe the chronological development of tools for EPIs, chromatin interactions/loops, and boundaries of TADs. Finally, we describe the advantages and disadvantages of each tool.

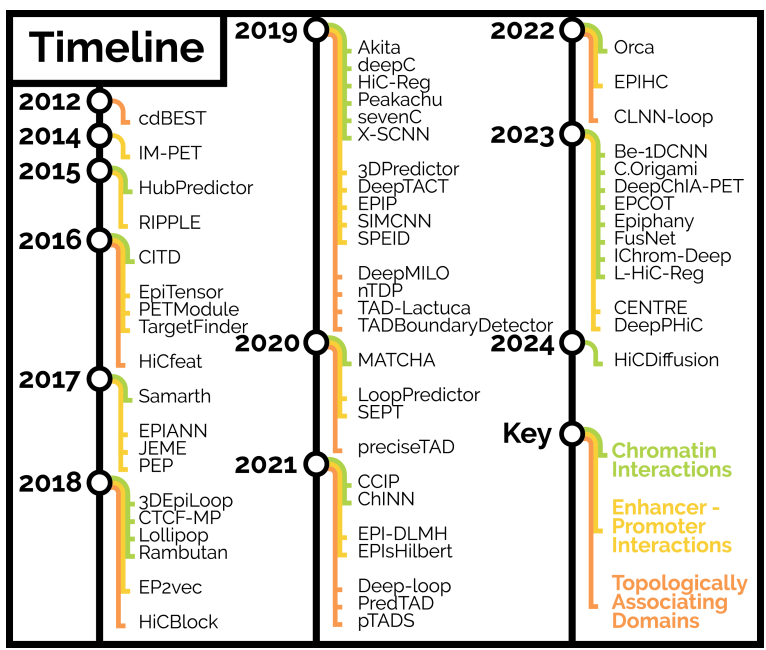

**Figure 1. Timeline of chromatin interaction prediction tools.**

## Results

**Machine learning framework for genomic predictions**

The most common approach for genomic predictions involves binary classification, such as distinguishing between functional and nonfunctional EPIs, boundaries and non-boundaries of loops, and TADs. Binary classification involves construction of predictor and response variables, also known as dependent and independent (explanatory) variables. A genome is binned into discrete regions and the dependent variable is constructed by either labeling those regions as being promoter, enhancer, or TAD/loop boundary (1), or not (0). The predictor variables are constructed from genome annotation data associated with binned regions. Various types of annotations can be used as predictors; DNA sequence/TFBS motifs, evolutionary conservation, overlap with TFBSs, and/or histone modification or open chromatin regions, etc. Each of these data require different computational approaches. For sequence-based features, one-hot encoding (DNA representation as a binary vector with four dimensions) or k-mer counts can be used. For genomic annotations, average or sum of signal within a region or the extent of overlap between a region and an annotation can be utilized [49]. A special case includes modeling pairs of regions (such as EPIs or two boundaries flanking a TAD) for which distance between these regions may be used as a response variable [30]. The proper construction of predictor and response variables is essential for genomic machine learning.

Genomics machine learning frequently suffers from several pitfalls leading to biased or overly optimistic results (reviewed in [50]). Class imbalance, defined as an unequal number of positive and negative examples, is one of such pitfalls. With a class imbalance, algorithms tend to focus on the class represented by the largest number of examples [51]. Methods to account for class imbalance include Random Over-Sampling (ROS), Random Under-Sampling (RUS), and combinations of the two (reviewed in [52]). ROS aims to balance class distribution through random replication of minority class examples. RUS performs random elimination of majority class examples. Another popular technique is the Synthetic Minority Over-sampling Technique (SMOTE). SMOTE forms new minority class examples by interpolation from K-Nearest Neighbors (KNN) of randomly selected minority class examples. Alternatively, proportionately up-weighting the misclassification cost for the minority class is another approach to account for class imbalance [53]. While some machine learning methods, such as Support Vector Machines (SVMs), can provide relatively robust classification results when applied to imbalanced data sets [54], deep neural networks are not immune to class imbalance, and methods like SMOTE can help [55,51]. Ultimately, the goal of data-level methods is to minimize class imbalance while making the data directly compatible with machine learning algorithms and performance evaluation metrics.

If class imbalance is not accounted for, common performance evaluation metrics may report overly optimistic performance. For instance, if the majority class has a 99% prevalence, an accuracy of 99% (or, equivalently, 1% error rate) can easily be achieved by always predicting the majority class. Such baseline methods can be notoriously difficult to beat [56]. The commonly used Area Under the Receiver Operator Characteristics curve (AUROC) metric [57] may demonstrate good performance, but the number of false positives could also be quite high because of the size of the majority class. Instead, Area Under Precision-Recall Curve (AUPRC) is better suited for performance evaluation as it assesses the fraction of true positives among positive predictions [58,59]. Alternatively, complementary evaluation metrics including combined metrics like F-measure, a harmonic mean of precision and recall [60], and Matthews Correlation Coefficient (MCC), a correlation coefficient utilizing all proportions of the predicted outcome [61], should be explored when working with imbalanced classes [62].

Multiple types of genome annotation data and multiple ways of constructing predictors are frequently used. Various techniques coupled with feature selection algorithms are frequently utilized to avoid overfitting. K-fold cross-validation randomly splits the data into *k* equal parts and uses *k* - 1 parts for training with the remaining part for validation. Leave-one-out cross-validation iteratively uses all but one data point for training with the remaining single data point for validation. Schreiber, J. et al. has described three cross-validation strategies, as well as their advantages and downsides [56]. The cross-chromosome and cross-cell-line validation techniques may lead to over-optimistic model performance as the average associations become memorized. In contrast, the hybrid approach minimizes the memorization effect by validating the model on unseen genomic locations (cross-chromosome) and in different data (cross-cell-line) [56]. Improper segregation of test and validation datasets may also lead to class imbalance and inflation of performance metrics [63]. We describe the aforementioned considerations for each tool where relevant.

**Using genome annotation data for prediction**

It has long been noted that different genomic annotations tend to co-localize, suggesting shared functional properties of a given region. Hidden Markov Model (HMM)-based methods have been successfully applied for segmentation of chromatin into distinct functional states. The HMM-SA method developed by Won, KJ. et al. [64] predicts human promoters and enhancers using signal from several histone modification marks shown to have complementary information for prediction. It was trained on RNAPII, TAF1, and p300 ChIP-seq data and could successfully predict promoters and enhancers in different cell lines. ChromHMM [65] and Segway [66] applies a HMM and its extension, a

Dynamic Bayesian Network (DBN), for learning chromatin states characterized by similar patterns of genomic annotations such as transcriptionally active or repressed regions, candidate enhancers, and promoters. The follow-up efforts annotate a genome in 20 annotation-based chromatin states (IDEAS [67]) and 100 conservation-based states (ConsHMM [68]).

Early methods demonstrating association between genome annotations and regulatory regions included prediction of promoters and enhancers from epigenomic data. CSI-ANN was among the first using artificial neural networks to predict enhancers [69]. Using six histone modification marks, the method implements two data transformation strategies, average signal and energy, for feature representation. Although the method uses a five-fold cross-validation scheme, the network is intentionally trained on an imbalanced dataset with the number of negative classes being 10X larger than positive classes. The empirically tuned time-delay neural network outperforms correlation-based and HMM-based methods for enhancer prediction as quantified by sensitivity and Positive Predictive Values (PPV). Consequently, it is applied to enhancer prediction in CD4+ cells, and is validated using p300 binding peaks, sequence conservation, and enrichment in TFBSs. CSI-ANN demonstrates that data transformation (energy transformation of epigenomic signal) may enhance predictive power of machine learning algorithms.

Machine learning methods, such as SVMs, are also adapted to predict enhancers from histone modification signal. ChromaGenSVM is an SVM method tuned with a genetic algorithm to select optimal parameters [70]. Using signal from six histone modification marks, it outperforms previous methods in predicting experimentally validated enhancers judged by the F1 score and AUROC metrics in cross-validation settings. This study also uses an imbalanced dataset with respect to the number of negative examples. Expectedly, H3K4me3 was the most predictive mark of enhancers, with H3K4ac and H2BK5ac marks adding to the predictive power.

RFECS [71] uses only genome annotation data to predict enhancers with a Random Forest (RF) algorithm. Trained on p300 sites using 5-fold cross-validation, the model identifies H3K4me1, H3K4me3, H3K4me2 marks as most predictive. Interestingly, the authors intentionally used imbalanced classes (1:7 to 1:9) arguing that when measured by AUROC, such models better reflect the observed excess of the negative class. They demonstrated that the model trained in one cell type can be used in another cell type, albeit at reduced validation and increased misclassification rates.

Besides genome annotation data, DNA sequence properties, such as k-mers, are also used to predict enhancers from DNA sequence [72]. The SVM model was trained on experimental enhancers using balanced, 50X imbalanced, and 100X imbalanced classes. Both AUROC and AUPRC were used for model evaluation in 3-fold cross-validation settings. The SVM model shows good performance in predicting enhancers vs. background sequence as well as distinguishing enhancers between tissues with the selected k-mers corresponding to TFBS motifs. CLARE [73] is another method for predicting active regulatory elements in a DNA sequence based on the presence of TFBS motifs. It uses Least Absolute Shrinkage and Selection Operator (LASSO) penalized regression to select a set of motifs best predicting active elements, as compared with matched-in-length and GC content random sequence. Although it uses sequence information only, the model achieved AUROC 0.83 in distinguishing forebrain enhancers from background sequences. Ultimately, gkm-SVM [74] and its improved version, LS-GKM [75], demonstrated good AUROC and AUPRC performance metrics.

Combining genome annotation data and DNA sequence information has also been explored for regulatory element prediction. EnhancerFinder [76] uses binding motifs together with evolutionary conservation and epigenomic data to predict enhancers using the SVM-based multiple kernel learning method. First, it distinguishes enhancers from non-enhancers, and second, predicts their activity in a given tissue. Trained on a balanced set of 711 experimentally validated VISTA enhancers and the same number of matched random regions, EnhancerFinder demonstrates that multiple datasets provide complementary information improving prediction accuracy compared to using a single predictor. Interestingly, both evolutionary conservation and DNA motifs were among the least predictive features, strengthening the notion that epigenomic data is better suited for enhancer prediction.

Numerous other methods have been developed for predicting TFBSs, DNA methylation status, and gene expression level from genome annotation data (reviewed in [77,78]). The success of using genome annotation data for predicting genomic features on a linear genome (1D predictions) prompted the development of methods for predicting higher-order chromatin interactions.

**Table 1. Enhancer-promoter interaction prediction tools.** CNN - Convolutional Neural Network, GBM - Gradient Boosting Machines, LSTM RF - Random Forest, RNN - Recurrent Neural Network, SVD - Singular Value Decomposition.

| | | | | |
|---|---|---|---|---|
| **Epigenomic data-based methods** | | | | |
| IM-PET | RF | http://tanlab4generegulation.org/IM-PET.html | Perl | [79] |

| **Epigenomic data-based methods** | | | | |
|---|---|---|---|---|
| EpiTensor | SVD | http://wanglab.ucsd.edu/star/EpiTensor/ | Matlab | [80] |
| RIPPLE | RF and group LASSO | http://pages.discovery.wisc.edu/~sroy/ripple/queryg.php | Matlab | [81] |
| PETModule | RF | https://hulab.ucf.edu/research/projects/PETModule/ | Python | [82] |
| TargetFinder | Boosted trees, multiple techniques | https://github.com/shwhalen/targetfinder | Python | [30] |
| JEME | RF, LASSO | https://github.com/yiplabcuhk/JEME | Java | [83] |
| EPIP | Decision trees | https://github.com/amlantalukder/EPIP | Python | [84] |
| 3DPredictor | GBM | https://github.com/labdevgen/3DPredictor | Python | [85] |
| LoopPredictor | RF, GBM | https://github.com/bioinfomaticsCSU/LoopPredictor | Python | [86] |
| **DNA sequence-based methods** | | | | |
| PEP | GBM | https://github.com/ma-compbio/PEP | Python | [87] |
| EPIANN | Attention-based DNN | https://github.com/wgmao/EPIANN | Python | [88] |
| SPEID | CNN+RNN | https://github.com/ma-compbio/SPEID | Python | [89] |
| SIMCNN | CNN | https://github.com/zzUMN/Combine-CNN-Enhancer-and-Promoters | Python | [90] |
| EP2vec | GBM | https://github.com/wanwenzeng/ep2vec | Python | [91] |
| DeepTACT | CNN-based | https://github.com/liwenran/DeepTACT | Python | [92] |
| SEPT | CNN+LSTM | https://github.com/NWPU-903PR/SEPT | Python | [93] |
| EPIsHilbert | CNN+Hilbert encoding | https://github.com/zmyqx/E-HilbertEPIs | Python | [94] |
| EPI-DLMH | CNN | https://github.com/Xzenglab/EPI-DLMH | Python | [95] |
| EPIHC | CNN+feed-forward+communicative module | https://github.com/BioMedicalBigDataMiningLab/EPIHC | Python | [96] |
| **Hybrid methods** | | | | |
| CENTRE | GBM | https://github.com/slrvv/CENTRE | R | [97] |
| DeepPHiC | CNN+shared feature extractor+classifier | https://github.com/lichen-lab/DeepPHiC | Python | [98] |

**Enhancer-promoter interactions prediction**

Enhancer-promoter interactions (EPIs) represent an easily understood functional link between distal genomic regions (enhancers) and gene promoters. An active EPI is thought to facilitate gene expression by recruiting transcriptional co-factors. Biological interpretability of EPIs prompted a plethora of tools being developed for their prediction (Table 1). Although not implemented as a tool, PreSTIGE was among the first methods to use epigenomic data for EPI prediction [99]. Considering cell type-specific H3K4me1 signal (a marker of enhancers), the method linked these regions with genes expressed specifically in that cell type. The method restricts its search to enhancers and genes located within domains marked by CTCF binding. The authors' estimation of the False Discovery Rate (FDR) using experimental 3C, 5C and ChIA-PET data demonstrated only ~40% of genes were regulated by the nearest enhancers with the majority of enhancers being located within 100Kb of transcription start sites.

He, B. et al. developed the IM-PET method to predict EPIs [79]. The IM-PET method estimates enhancer activity using the precomputed histone and transcription factor signal scores together with FPKM gene expression between enhancers and promoters. The method additionally includes both coevolution and distance between enhancers and promoters with distance proving to be the most predictive. IM-PET was trained with a RF model on a balanced dataset using ChIA-PET-validated positive EPIs. The model outperformed a baseline nearest promoter metric, SVM and logistic regression, and PreSTIGE. Five-fold cross-validation, AUROC, and F1 scores were used to assess performance. The authors predicted EPIs in 12 cell lines and reported that one enhancer targets about three promoters on average.

EpiTensor [80] predicts cell type-specific EPIs using singular value decomposition of a tensor comprised of signal measures of 16 histone marks, DNase Hypersensitive Sites (DHSs), and RNA-seq signals in five cell types. it decomposes the tensor into three subspaces (cell-type, assay, and genomic locus subspaces), focusing on the "locus" eigenvectors. The genomic locus subspace captures epigenomic patterns of interacting distal regions and demonstrates the importance of H3K4me1, H3K4me3, H3K27ac, and DHSs in defining EPI hotspots. Although not considering the class imbalance problem, the method outperforms the correlation-based baseline as measured by AUROC. EpiTensor has enabled prediction of EPIs at a 200bp resolution.

Roy, S. et al. developed RIPPLE [81], a method using a combination of RFs and group LASSO to predict enhancer-promoter interactions in multiple cell lines. It uses 3C chromatin interaction data with 23 epigenomic datasets (8 histone marks, 15 TFBSs), common for five cell lines. The method addresses class imbalance with RUS controlling for distance, but the model performs well even on imbalanced data as judged by AUPR. CTCF, SMC3, RAD21, DNase I, expression, H3K27ac, H3K27me3, H3K36me3, H3K4me2, H3K4me3, H3K79me2, H3K9ac, RNA PolII and RAD21 were the most predictive and the model trained in one cell line could predict EPIs in anther.

Zhao, C. et al. developed PETModule [82], a RF based approach for predicting EPIs. It uses conservation, distance, enhancer-promoter gene ontology correlation, and a novel sequence motif module feature. PETModule's feature selection identifies distance as the most predictive feature. Compared with IM-PET and PreSTIGE, the method has better recall, precision, AUROC, and F1 score when trained on balanced data and validated on experimental and manually curated interactions. The model was trained on human datasets and performs well in mouse datasets. The authors confirmed that the majority of enhancers are distant and should be considered within 2Mb regions from transcription start sites.

The popular TargetFinder [30] method uses boosted trees to predict EPIs using DNase-seq, DNA methylation, TF ChIP-seq, histone marks, CAGE, and gene expression data. Similar to IM-PET, the authors identified the window feature (the intervening genomic interval between the enhancer and the promoter) as the most predictive. Interestingly, the best performance with an F1-score of approximately 0.2 was only achieved by gradient boosting with a high number (~4000) of trees but not with a linear SVM. The follow-up investigation of these results demonstrated that a class imbalance in the test set with respect to the number of enhancers overlapping a promoter leads to overly optimistic prediction performance. Consequently, cross-validation on a sorted chromosome list (training on one and testing on the following chromosome) demonstrates nearly random prediction performance [63].

JEME [100] uses a two-step process to define enhancer promoter interactions. First, functional enhancers are identified using a LASSO regression of gene expression on signal from histone modifications and DHSs within 1Mb across samples. Second, to predict the regulating enhancers in a particular sample, it uses a RF model constructed on sample-specific error terms and features, including distance. RUS is used to address class imbalance in cross-validation settings and performance assessment using AUPRC. The authors show that the method outperforms IM-PET, PreSTIGE, Ripple, and TargetFinder, and predicted EPIs for 935 cell and tissue types.

Enhancer-Promoter Interaction Prediction (EPIP) [84] was pioneering when it was introduced by combining genomic (distance, conservation synteny score, and correlation of epigenomic signals in enhancers and promoters) and epigenomic data (DNAse, TFBS, and histone ChIP-seq). It is an ensemble learning method using 200 weak learners (decision trees) and an AdaBoost-inspired algorithm to reweight the input data after each learner. The positive and negative classes were created from FANTOM enhancers, subsetted by cell type-specific ChromHMM enhancer annotations, and loops detected from Hi-C data. The model was trained on both balanced and imbalanced data based on the observation that the balanced model had a high sensitivity, and the imbalanced model had a high specificity when tested on the training data by 10-fold cross-validation. The authors show the model performed well in cells with incomplete or missing data and outperformed TargetFinder and Ripple as judged by AUROC, AUPRC, and F1 scores.

Belokopytova, P. S. et al. [85] developed 3DPredictor, a gradient boosting model for predicting EPIs using CTCF binding loci and orientation, gene expression, and distance between interacting loci. The authors showed the inferior performance of TargetFinder was due to the incorrect design of training and validation datasets and developed a training strategy using balanced classes with a cross-chromosomal training-validation approach and multiple evaluation metrics to overcome its limitations. Instead of predicting qualitative EPIs, they predict the quantitative strength of EPIs. They showed the model trained on one cell type can successfully predict EPIs in another cell type.

Tang, L. et al. developed LoopPredictor [86], a two-component ensemble machine learning model to predict EPIs and chromatin loops. The model architecture included an Anchor type predictor implemented with a hybrid RF and Confidence predictor using Gradient Boosted Regression Trees. The model uses TFBSs, histone modifications, chromatin accessibility, gene expression, methylation, and transcription start sites to train the model on H3K27ac and YY1 HiChIP data. The authors showed that the model outperforms the TargetFinder and 3DPredictor tools as well as four baseline classifiers (Linear Support Vector Clustering (SVC), Logistic Regression, K-Neighbors, and RF) as judged by the F1 score, AUROC, and AUPRC in cross-validation settings. as well as TargetFinder and 3DPredictor tools. Trained on human data, the model performed well in predicting EPIs across other organisms. The authors demonstrated, not single, but multiple genomic feature types were needed for accurate predictions. The continued demonstration of the superiority of the RF technique over other methods [101] positions it as a robust machine learning approach for EPI predictions utilizing epigenomic data.

It has also been demonstrated that tools can predict EPIs using only DNA sequence. PEP [87] uses a machine learning model and Gradient Tree Boosting [102] based only on features from DNA sequences of enhancer and promoter regions. It has two variants, PEP-Motif and PEP-Word. PEP-Motif only uses motif enrichment features for known TFBS motifs. PEP-Word uses word embeddings, a term from natural language processing, that allow representing (arbitrary-length) sentences from a discrete vocabulary as fixed-length numerical vectors while retaining semantic meaning in the embedding space. The authors addressed class imbalance by reweighting and using multiple performance metrics (AUROC, AUPRC, Precision, Recall, F1, MCC). The authors' results show that it is possible to achieve comparable results using only sequence-based features to predict EPIs.

Similarly, EPIANN [88], an attention-based neural network model, was developed for predicting EPIs with only DNA sequence. The network structure includes three functional blocks; an attention mechanism, interaction quantification, and multi-task learning block. EPIANN was trained on TargetFinder's data and accurately predicted EPI events. It also reveals pairwise attention scores which uncovers over-represented TFBSs and TF-pair interactions associated with enhancer function. Comparative analysis with TargetFinder and PEP shows comparable performance across cell lines, with EPIANN outperforming PEP based on AUROC scores while demonstrating similar performance to TargetFinder in other scenarios. Attention-based analysis provides insight into the importance of DNA sequence features and correlates well with known genome annotations like CTCF and EGR family members. Attention-based analysis also identifies novel TFs driving EPI formation like NRF1 for which ChIP-seq data is not available. The model's multi-task learning architecture ultimately enhances generalizable feature learning, while motif analysis and DNAseq footprinting improve biological interpretation.

The authors of PEP also developed SPEID [89], a deep neural network framework predicting EPIs from DNA sequence only. Its architecture includes combining convolution, activation, and max-pool layers, together with a recurrent neural network (RNN) layer and a dense layer. Class imbalance was addressed using a "data augmentation" method similar to oversampling. Several metrics were used to benchmark performance (AUROC, AUPRC, F1) and the method outperformed PEP and TargenFinder. Zhuang, Z. et al. [90] proposed a simplified version of a deep neural network used by SPEID, without the RNN layer (SIMCNN). They demonstrated the network's performance to be comparable or better to that of SPEID, TargetFinder, EPIANN, and PEP. Similarly trained and evaluated, their method performed well in cell type-specific settings but less optimally across cell lines.

EP2vec [91] uses sequence embedding features of enhancers and promoters to train a Gradient Boosting Regression Trees algorithm on experimentally validated EPI pairs. EP2vec uses balanced classes, with the same number of positive and negative examples, and evaluated using F1, AUROC, AURPC in 10-fold cross-validation settings. Although it outperformed the epigenomic data-based TargetFinder and SVM predictors, the addition of epigenomic data slightly improved performance, suggesting that sequence and epigenomic features are complementary.

The DeepTACT neural network was developed to predict EPIs and promoter-promoter interactions from DNA sequence and chromatin accessibility (DNAse-seq) data [92]. It contains three modules: a sequencing module using one-hot encoded sequence information, an openness module using chromatin accessibility signal, and an integration module that includes a bidirectional Long Short-Term Memory (LSTM) network and an attention layer. The authors showed DeepTACT's use of two data types improved performance as measured by AUROC and AUPRC. Due to multi-way interactions, they justified the use of imbalanced data. DeepTACT's data augmentation and bootstrapping strategy was designed to improve the stability of network parameters trained on experimental Promoter-Capture Hi-C and ChIA-PET data. Compared with SPEID and Rambutan, DeepTACT detects interactions at finer resolution, is more biologically interpretable, and discovers hub promoters active across cell types. Similar to EP2vec, this method demonstrates that DNA sequence properties complement epigenomic features for functional element predictions.

Jing, M. et al. developed SEPT [93], a deep learning method using DNA sequence information for EPI prediction. It extracts sequence features of enhancers and promoters from DNA sequence using two CNN layers and one LSTM layer. Then, it introduces the gradient reversal layer to reduce the cell line specific features and prioritize enhancer-promoter specific features. SEPT simultaneously trains two classifiers of EPIs and domain-specific (cell type specific) discriminators. Trained on cell lines with known EPIs, the authors showed that SEPT performed well in new cell lines and outperformed LS-SVM, SPEID, and RIPPLE according to AUROC, AUPRC, and F1 scores.

Zhang, M. et al. introduced EPIsHilbert [94], a convolutional neural network (CNN) utilizing Hilbert curve encoding to predict EPIs. Hilbert curve encoding of DNA sequence and improved the model's performance by preserving the spatial positioning relationship between an enhancer and a promoter. The authors addressed class imbalance using both over and under sampling techniques and used F1, AUROC, and AUPRC metrics for performance evaluation.

Their approach outperformed SPEID and SIMCNN using different cell line data. This work introduces additional approaches for EPI prediction, such as transfer learning between cell lines and visualization of sequence features.

Min, X. et al. developed EPI-DLMH [95], a two-layer CNN for predicting EPIs from only DNA sequence. EPI-DLMH contains a bidirectional, gated, recurrent unit to capture long-range dependencies, and an attention mechanism to prioritize the most important features. It also uses k-mer sequence representation and embedding with dna2vec to reduce dimensionality. The authors additionally introduced an additional matching heuristic to capture more interaction information between promoters and enhancers. Using balanced cell type-specific datasets and several performance evaluation metrics (AUROC, AUPRC, F1), their method outperformed SPEID and EPIANN. This approach demonstrates that additional DNA sequence information captured by matching heuristics can improve EPI prediction.

The EPIHC [96] deep neural network is another method combining sequence-derived and genomic annotations. It extracts sequence features from enhancers and promoters using CNN. It then introduces a communicative learning module to capture communicative information between enhancers and promoters. The authors investigated the effect of class imbalance on the model's performance in 5-fold cross-validation settings using AUROC and AUPRC. Using benchmarking data from TargetFinder, the model was trained on data from multiple cell lines and then applied to a target cell line. EPIHC outperforms three other neural network-based methods, SIMCNN, SPEID, and EPIVAN. This paper demonstrated that hybrid data (both sequence and genome annotation data) and model usage (neural networks connected via a communicative learning module) improves EPIs.

Combining epigenetic data with other data modalities, such as gene expression, showed to be beneficial for EPI predictions. Cell-specific ENhancer Target pREdiction (CENTRE) [97] aims to predict cell type-specific EPIs using only gene expression and ChIP-seq data (H3K27ac, H3K4me1, and H3K4me3). It trains the XGBoost algorithm on the ENCODE registry of candidate cis-regulatory elements with enhancer-like signatures and GENCODE transcription start sites. It also utilizes the Benchmark of candidate ENhancer-Gene Interactions (BENGI) dataset for validation to mitigate bias stemming from dependencies between training and test datasets. The authors used 12 cross-validation groups divided by chromosome and evaluated performance with F1-scores due to the highly imbalanced dataset. The model outperformed TargetFinder and, similar to IM-PET, PETModule, and other methods, identified enhancer-promoter distance as the most predictive feature.

DeepPHiC [98] is a multi-modal deep learning model for EPI and promoter-promoter interaction prediction. It incorporates one-hot encoded DNA sequence, epigenetic signals (H3K4me1, H3K4me3, H3K27ac), anchor distance, evolutionary, and DNA structural features. The model's architecture, based on DenseNet, comprises two modules: a feature extractor (two dense blocks and a CNN) and a classifier. It was trained on promoter-centered interactions from the 3D-genome Interaction Viewer and database using an unbalanced dataset (5x negative examples). DeepPHiC outperforms SPEID, DeepMILO, ChINN, and DeepTACT as evaluated by AUROC and AUPRC. This approach demonstrates superiority in predicting EPIs and promoter-promoter interactions across multiple cell types using a combination of genomic and epigenomic features with distance between anchors.

**Table 2. Chromatin interaction prediction tools.**

| Epigenomic data-based methods | | | | |
|---|---|---|---|---|
| HubPredictor | Bayesian Additive Regression Trees | https://github.com/huangjialiangcn/HubPredictor | R | [103] |
| CITD | Wavelet deconvolution, non-linear transformation | https://cb.utdallas.edu/CITD/index.htm | NA | [104] |
| 3DEpiLoop | RF | https://bitbucket.org/4dnucleome/3depiloop | R | [105] |
| Lollipop | RF | https://github.com/ykai16/Lollipop | Python | [106] |
| HiC-Reg | RF regression | https://github.com/Roy-lab/HiC-Reg | C | [107] |
| L-HiC-Reg | RF regression | https://github.com/Roy-lab/Roadmap_RegulatoryVariation | C | [108] |
| sevenC | Logistic regression | https://bioconductor.org/packages/release/bioc/html/sevenC.html | R | [109] |
| Epiphany | CNN+LSTM | https://github.com/arnavmdas/epiphany | Python | [110] |
| X-SCNN | Siamese CNN | https://github.com/ernstlab/X-CNN | Python | [111] |

| **Epigenomic data-based methods** | | | | |
|---|---|---|---|---|
| EPCOT | Encoder-decoder | https://github.com/liu-bioinfo-lab/EPCOT | Python | [112] |
| **DNA sequence-based methods** | | | | |
| Samarth | SVM | https://bioinf.mpi-inf.mpg.de/publications/samarth/ | Matlab | [113] |
| Akita | CNN | https://github.com/calico/basenji/tree/tf2_hic/manuscripts/akita | Python | [114] |
| deepC | CNN | https://github.com/rschwess/deepC | Python | [115] |
| ChINN | CNN | https://github.com/caofan/chinn | Python | [116] |
| Orca | Encoder-decoder | https://github.com/jzhoulab/orca | Python | [117] |
| HiCDiffusion | Encoder-decoder, CNN, transformer | https://github.com/SFGLab/HiCDiffusion | Python | [118] |
| **Hybrid methods** | | | | |
| CTCF-MP | Boosted Tree Classifier | https://github.com/ma-compbio/CTCF-MP | Python | [119] |
| Rambutan | CNN | https://github.com/jmschrei/rambutan | Python | [120] |
| CCIP | RF | https://github.com/GaoLabXDU/CCIP | Python | [121] |
| IChrom-Deep | Transformer | https://github.com/HaoWuLab-Bioinformatics/IChrom-Deep | Python | [122] |
| FusNet | Fusion network | https://github.com/CSUBioGroup/FusNet | Python | [123] |
| C.Origami | CNN, transformer | https://github.com/tanjimin/C.Origami | Python | [124] |
| **Hi-C data-based methods** | | | | |
| Peakachu | Random Forest | https://github.com/tariks/peakachu | Python | [125] |
| MATCHA | Graph embedding+NN | https://github.com/ma-compbio/MATCHA | Python | [126] |
| Be-1DCNN | Bagging of 1D CNN models | https://github.com/HaoWuLab-Bioinformatics/Be1DCNN | Python | [127] |
| DeepChIA-PET | CNN | https://github.com/zwang-bioinformatics/DeepChIA-PET/ | Python | [128] |

**Chromatin interaction prediction**

Besides canonical EPIs, predicting significant chromatin interactions using genome annotation and/or DNA sequence data also attracted significant attention (Table 2). Similar to EPIs, these interactions are thought to facilitate long-range gene expression regulation and are frequently referred to as chromatin loops, interaction hubs, or significant chromatin interactions. Huang, J. et al. 2015 [103] developed HubPredictor that predicts frequently interacting genomic loci (termed "hubs") and TAD boundaries from nine histone modification marks and CTCF binding sites. They employed a Bayesian Additive Regression Trees (BART) model that used signal from CTCF and nine cell-type-specific histone modification marks. The model was trained on balanced data (RUS), evaluated using cross-validation and AUROC, and performed well across cell lines. Histone marks (H3K4me1) were the most predictive of chromatin hubs across datasets and cell types. The addition of sequence-based features, such as conservation and GC content, did not improve predictions. TAD boundaries were predicted by CTCF peaks and negative H3K4me1 peaks. Alongside this work, Dixon, J. R. et al. [25] predicted differential chromatin interactions using a RF algorithm. It was trained on the signal measures of six histone modification marks, CTCF, and DNAse hypersensitive sites on balanced data using RUS. Similarly identified changes in H3K4me1 signal were the most predictive followed by DHSs.

Chen, Y. et al. [104] developed CITD, a method to predict genome-wide chromatin interaction matrices using 1D histone modification data. The authors utilized the observation that histone modification signals in nearby genomic bins are correlated, and this correlation follows power-law decrease with the increasing distance between bins. They developed a five-step procedure that includes wavelet decomposition of the original Hi-C matrix, nonlinear power-law transformation of the resulting coefficient matrices, and wavelet deconstruction of the predicted interaction matrix.

Both cross-chromosome and cross-cell-line validation showed high correlation of predicted and experimental matrices, similar TADs, and EPIs.

Al Bkhetan and Plewczinski [105] developed 3DEpiLoop to predict two types of chromatin loops: CTCF-mediated and RNAPII-mediated. They trained a RF model in genome-wide cross-validation settings and estimated the performance of their model with both AUROC and experimentally obtained ChIA-PET data as a gold standard. In their tests, the RF technique outperformed AdaBoost classification trees, neural networks, SVM, and Stochastic Gradient Boosting. This study was among the first that used distance between genome annotations and loop anchors. The authors showed differences in genome annotation signatures with CTCF-mediated loops best predicted by TFBSs and RNAPII-mediated loops best predicted by histone marks, which indicated differences in annotation signatures for different loop types.

A similar approach was taken by Kai, Y. et al. [106], who developed Lollipop to predict CTCF-mediated interactions using a RF technique. Besides TFBS signal and orientation, they used conservation scores and the distance between loop anchors (loop length). Trained on experimental ChIA-PET data, the model was evaluated in cross-validation settings using both AUROC and AUPRC. Similar to previous models, the authors identified CTCF, RAD21, and loop length as the most predictive features. Models trained in one cell line performed well in another cell line and also identified many more loops than experimentally detected. This work demonstrated that genome annotations can guide the discovery of novel loops that cannot be detected by current 3C technologies.

Zhang, S. et al. developed HiC-Reg [107], a RF regression-based approach for predicting genome-wide chromatin interactions from epigenomic data. The authors experimented with various feature encoding approaches to predict interactions between two regions. These approaches include concatenating epigenomic signals at both regions, including signal between regions, and incorporating epigenomic signals from multiple cell lines. Using the model trained on the distance between interacting regions as a baseline, they demonstrated improved performance when including signal between regions and across cell lines according to the AUROC scores. They used cross-chromosomal validation within and between cell lines with an expected drop in performance when using chromosomes and cell lines other than those used for training. The authors demonstrated the biological relevance of their predictions using distance-stratified Pearson Correlation Coefficients (PCCs) between the original and predicted interactions. They detected similar regions forming loops, bidirectional CTCF binding in those loops, and overlapping TAD boundaries. Using experimental evidence of interactions in HBA1 and PAPPA gene promoters, they demonstrated their model can successfully predict them. An extension of this model, L-HiC-Reg, performs local modeling and prediction of chromosomal interactions, and additionally evaluates networks of significant gene interactions and the associated disease-associated variants [108].

The sevenC R package by Ibn-Salem & Andrade-Navarro [109] investigated the hypothesis that a functional chromatin loop interaction should contain highly correlated CTCF ChIP-seq signal. Consequently, a model trained on such correlation measures may be used to predict functional chromatin loops. They considered every pair of CTCF motifs in a convergent orientation within a 1Mb window as potential looping interactions. For each pair, they measured the correlation of ChIP-seq signal within a 1000bp window around each motif. Using the correlation coefficient, distance between motifs, orientation, and motif significance scores, they built a logistic regression model using high-resolution Hi-C and ChIA-PET data as a gold standard. They acknowledged the presence of class imbalance, although did not directly account for it, and they used multiple performance evaluation metrics (AUROC, AUPRC, F1-score). Their approach revealed comparable predictive power using CTCF, RAD21, and ZNF143 factors, known architectural proteins of chromatin loops [129,8], as well as novel factors TRIM22, RUNX3, and BHLHE40. Interestingly, DNAse-seq performed similarly to TF ChIP-seq signal; however, histone modification signal (H3K4me1, H3K4me3, H3K27me3, H3K27ac) was not predictive. Their combined model, trained on the average signal of the 10 most predictive TFs, can be used to predict cell type-specific loops.

Deep neural network architectures were also explored for chromatin interaction predictions. Epiphany [110] was among the first neural networks to predict the Hi-C contact map from five epigenomic tracks (DNAse, CTCF, H3K27ac, H3K27me3, and H3K4me3). As an architecture, it uses 1D convolutional layers and a bidirectional LSTM. The authors further used a generator network to extract information and make predictions and a discriminator to use adversarial loss to predict realistic Hi-C maps. AUROC and AUPRC were used for performance evaluation. The authors demonstrated that CTCF was critically important for accurate Hi-C map prediction and its removal breaks the model.

EpiMCI [130], although not implemented as a tool, uses a dual-channel hypergraph neural network for reconstructing multi-way chromatin interactions from epigenomic signals. The model represents the data as a hypergraph, learns the vertex embedding and predicts multi-way interactions as a classification task. It was developed for high-throughput Pore-C technology and applied to GM12878 and K562 cells at 1Mb distance. It uses balanced classes, cross-validation, and six evaluation metrics. It marginally outperformed MATCHA, described below, in predicting multi-way chromatin interactions. Besides prediction, embeddings can be used to define A/B compartments and denoise Pore-C data.

EPCOT [112] is a deep learning framework (an encoder-decoder architecture) for predicting cell-type specific chromatin interactions, as well as epigenomic data, gene expression, and enhancer activity from chromatin accessibility only (DNAse-seq, ATAC-seq). It utilizes a pre-training and fine-tuning approach, where a model is first trained on cell-specific chromatin accessibility profiles (one-hot encoded accessibility sequences) and then transferred to downstream tasks. EPCOT's performance is rigorously validated through several experiments, including cross-chromosome and cross-cell type prediction analyses, where it consistently outperforms baseline models, such as Avocado for epigenomic data prediction, as evidenced by AUROC and AUPRC. Additionally, the authors demonstrate the model's ability to accurately predict TF binding patterns and its generalizability across different cell types.

Jaroszevicz & Ernst developed X-SCNN [111], a siamese CNN (two subnetworks with shared parameters) that leverages signal from TFBSs, histone marks, DNAse data (average signal) to predict chromatin interactions at high resolution (100bp). They used HiCCUPS-called interactions to train the model, RUS to balance the data, and used AUROC, AUPRC, and chromosome-specific validation to benchmark performance.

Besides genome annotation data, a polymer physics model [131] has been proposed to simulate chromatin interactions at high resolution. The model uses 15 chromatin states learned from histone modification profiles, CTCF motifs and their orientation, and employs molecular dynamics simulation based on an energy function to simulate an ensemble of high-resolution chromatin structures. Using a variety of similarity metrics (PCC, stratum-adjusted correlation, contact enrichment, and TAD boundary matching score), the model demonstrates the feasibility to resolve small chromatin loops, TADs, and long-range EPIs compatible with Hi-C and imaging data.

Similar to EPI predictions, DNA sequence properties have also been explored for chromatin interaction predictions. Nikumbh & Pfeifer developed Samarth [113], a tool for predicting long-range chromatin interactions using sequence information only. They used SVM with string kernel predictors using the oligomer distance histogram representations (histogram of distances between short (3-5bp) oligomers in the sequence) and trained their model on significantly interacting vs. non-interacting loci. The model accounts for class imbalance using upweighting of misclassification cost and uses AUROC for performance evaluation. This work demonstrated that short tandem repeats are potentially important for distinguishing interaction patterns. Although the model's performance was relatively modest in predicting cell type-specific interactions, this work was among the first to demonstrate that sequence alone provides information for long-range chromatin interaction predictions.

Fudenberg, G. et al. developed Akita [114], a CNN built using the Basenji architecture [132] that processes 1Mb DNA sequences and predicts chromatin interactions at 2Kb resolution. Akita was trained on high-resolution human Micro-C data and accounts for distance-dependent decay of interactions in chromatin interaction data. A conceptually similar CNN, deepC [115], similarly predicts chromatin interactions from DNA sequence, but uses epigenomic features to pre-train and initialize the network. It is trained on cell-type-specific Hi-C datasets validated using high-resolution Capture-C interactions, and it was shown to outperform HiC-Reg. These tools allow for understanding the effect of mutations (in silico saturation mutagenesis), understanding structural variants, interpreting expression quantitative trait loci (eQTLs), and predicting chromatin interactions in different species.

Cao, F. et al. developed ChINN [116], a CNN for predicting chromatin interactions from DNA sequence. The authors first showed that interactions can be predicted from both functional genomic data and distance between interacting regions. The CNN was then applied to forward and reverse complement sequences of interaction anchors and trained on CTCF loops, RNA PolII-mediated loops, and Hi-C data. The model was trained on a 1:5 imbalanced dataset and evaluated using AUPRC. The authors showed that convergent CTCF orientation is an important predictor, while other motifs complement predictive power.

Orca [117], a deep-learning model, predicts 3D genome architecture directly from genomic sequence data. Its architecture includes a hierarchical sequence encoder and a multilevel cascading decoder designed to provide a 'zooming' series of predictions at multiple scales. Performance evaluations show strong correlations between predicted and experimental interactions, with validation techniques such as AUROC scores confirming the model's efficacy. Orca accurately predicts diverse genome interaction mechanisms, including those mediated by CTCF and Polycomb, and exhibits concordance with experimental observations of chromatin interactions marked by histone modifications like H3K4me3, H3K27ac, and H3K4me1. Similar to other models for predicting chromatin interactions from DNA sequence, Orca enables in silico screens to probe sequence-based mechanisms of genome organization.

Recent deep learning approaches have also been utilized to predict chromatin interactions. The HiCDiffusion model [118] was developed to improve the resolution of Hi-C matrices generated from DNA sequence data. By incorporating an encoder-decoder architecture and a diffusion model, HiCDiffusion aims to reduce artificial blurring and enhance the fidelity of predictions. The encoder-decoder architecture includes residual convolutions and transformer encoders, with transfer learning utilized to pre-train the encoder-decoder architecture. Evaluation involved comprehensive data processing, training, testing, and validation procedures (Fréchet inception distance scores, used for quantifying the realism and diversity of generated images, correlation with the original Hi-C maps), with comparisons to C.Origami showcasing superior performance in both sequence-only and epigenetics-enhanced scenarios. These results underscore the potential of modern deep learning architectures for improving predictions of chromatin interactions solely from DNA sequence data.

A combination of genomic annotations and sequence features was also explored for chromatin interaction predictions. Zhang, R. et al. [119] developed CTCF-MP, a boosted tree classifier for predicting chromatin loops using CTCF motifs, distance between them, conservation, and cell type-specific ChIP-seq and DNAse-seq data. The CTCF motif sequences plus flanking regions were initially processed using word2vec and deep autoencoder to compress the 200-dimensional space to 32 dimensions, and the learned features were used for prediction. Imbalanced data performed well in their settings, with AUROC and AUPRC used for performance evaluation in cross-validation settings. The model trained on one cell type can be used to predict chromatin loops in other cell types, although the best performance was achieved on predicting common loops.

Schreiber, J. et al. developed Rambutan [120], a deep convolutional network for predicting significant chromatin interactions from one-hot encoded DNA sequence and DNAse Hypersensitive sites (signal, logFC over control). The network is trained on significant interactions defined by Fit-Hi-C (q-value <= 1e-6), DNAse signal, and binarized genomic distance. It uses balanced classes, AUROC, and AUPRC for performance evaluation. Predictions correlated with Insulation Score and replication timing. Rambutan is trained at a 1-5Kb resolution, but theoretically can predict at a finer resolution.

Network/graph analysis methods were also applied for chromatin interaction detection. Wang, W. et al. developed CTCF-mediated Chromatin Interaction Prediction (CCIP) [121], a tool for predicting CTCF-mediated convergent loops and tandem loops with transitivity. Transitivity is defined from the network of multiple CTCF-interacting regions, convergent and tandem. In addition to CTCF, RAD21 binding sites, anchor and in between features (as in TargetFinder), and directional CTCF motif one-hot encoding, the model incorporated the graph connecting probability (GCP) score which proved to be the most important predictive feature. A RF was trained on a balanced dataset and evaluated in cross-validation settings with AUROC, AUPRC, and other metrics. The authors showed that CCIP outperforms Lollipop and CTCF-MP and that transitive loops can explain the formation of tandem loops.

Farré, P. et al. [133] demonstrated the use of a deep neural network (dense architecture) to predict chromatin interactions using sequences of TFBSs (ChIP-seq) and transcriptionally active/inactive (RNA-seq) data. They also demonstrated that chromatin interactions themselves can be used to predict TFBSs. The following research utilized a CNN architecture to predict chromatin interactions with only DNA sequence.

The IChrom-Deep [122] neural network combines the use of sequence features and genomic features. It was implemented using a novel attention-based deep learning model containing a sequence module and a genomic module. The genomic module processes genomic features, conservation scores, CTCF motifs, and the distance between chromatin bins. The model was trained on a balanced dataset and evaluated using four evaluation metrics (AUPRC, Accuracy, MCC, F1) alongside a cross-validation strategy. The authors showed that IChrom-Deep outperforms three models, TargetFinder, XGBoolst and SGDC and demonstrated the importance of the distance of interacting regions, CTCF (convergent orientation), and the cohesin complex members (RAD21, SMC3). This tool demonstrates that synergistic use of sequence and genomic features can improve chromatin interaction predictions.

FusNet [123] is a 3-layer fusion neural network designed for predicting chromatin loops utilizing genome sequence information, distance between anchors, open chromatin, and ChIP-seq data. The feature extraction layer employs a CNN for dimensionality reduction of one-hot encoded DNA sequences. The predictor layer integrates Light Gradient Boosting Machine eXtreme Gradient Boosting and KNN models. The fusion layer integrates the predictions from each basic model as new features for model training and prediction of the final probability of loop formation. The fusion layer improves prediction performance, especially in cross-cell type loop prediction, as compared with ChINN, Peakachu, DeepYY1 methods using AUROC and AUPRC. FusNet demonstrates high consistency with Hi-C data and associates predicted loops with regulatory functions and disease-related mechanisms. Additionally, FusNet's prediction accuracy is supported by Aggregate Peak Analysis and aligns well with known TADs and EPIs. Permutation importance analysis highlights the significance of sequence features, particularly anchor sequences. FusNet was also applied to identify potential target genes of pathogenic single nucleotide polymorphisms (SNPs).

Similarly, C.Origami [124] uses DNA sequence, CTCF-binding, and ATAC-seq signals to predict genome organization comparable to high-quality Hi-C experiments. It introduces a novel multimodal deep neural network architecture for cell type-specific prediction of Hi-C maps in 2Mb windows. One-hot-encoded DNA sequence and feature signals are processed in parallel by 12 1D CNN layers followed by a transformer with eight self-attention blocks. The following 2D convolutional layers reconstruct the predicted Hi-C map. To evaluate the performance of C.Origami and other comparable models (Akita, deepC, Orca), the study uses insulation score correlation, observed/expected Hi-C map correlation, mean squared error (MSE), distance-stratified correlation, and AUROC. C.Origami and similar models facilitate in silico genetic perturbation studies, enabling efficient exploration of causal relationships in chromatin organization.

Hi-C data itself may contain sufficient information for chromatin interaction predictions. Salameh, T. J. et al. developed Peakachu [125], a RF model to predict chromatin loops (represented as pixels of high intensity) using Hi-C contact matrices only. They designed a strategy to represent each loop as a collection of intensities and ranks within a 11x11 pixel window around each loop. Using loops detected by H3K27ac, HiChIP, and CTCF ChIA-PET as positive examples with the same number of negative examples, they achieved robust performance measured by MCC. Compared with HiCCUPS, Peakachu detected more experimentally validated loops. The model shows robust performance when varying sequencing depth across cell lines and can be applied to data obtained by other technologies such as Micro-C and SPRITE.

MATCHA [126] uses structural information derived from Hi-C data and graph embedding followed by Hyper-SAGNN (a self-attention based graph neural network for hypergraphs) analysis. Multi-way chromatin interactions are represented as hyperedges that can be predicted by Hyper-SAGNN. Applied to SPRITE and ChIA-Drop data with Gm12878 at both 1Mb and 100Kb resolutions, the method provides a wealth of information about the genomic properties of multi-way interactions. This method was the first to predict multi-way chromatin interactions.

The Be-1DCNN model [127] employs a bagging ensemble of ten one-dimensional CNNs (each containing three convolutional layers and dropout layers) to improve prediction accuracy of chromatin loops. It is trained on high-resolution Hi-C data using 22 chromosomes for training and one for testing. Comparisons with other models like Gaussian Naïve Bayes, Perceptron, KNN, Decision Tree, and Peakachu showed Be-1DCNN's superior performance in predicting chromatin loops as evaluated on experimental ChIA-PET and HiChIP data. The model was trained on balanced datasets (RUS-like of negative examples) and its effectiveness is validated using various metrics such as accuracy, MCC, and area under the curve. This method demonstrated that combining traditional machine learning approaches, such as bagging, with deep learning architectures improves chromatin loop prediction.

Combining Hi-C and genomic annotation data was shown to be beneficial for chromatin interaction predictions. Liu, T. et al. presented DeepChIA-PET [128], a CNN (40 dilated residual convolutional blocks) for predicting ChIA-PET-defined chromatin interactions from Hi-C and ChIP-seq data. The model outperforms Peakachu when trained on Hi-C data only using three evaluation metrics (average precision, AUROC, AUPRC) with CTCF ChIA-PET representing ground truth. The model was trained on one cell line (GM12878) can be applied to others (e.g., HeLa). The authors demonstrated that performance improves when including ChIP-seq data, underscoring the importance of epigenomic data for chromatin interaction prediction.

**Table 3. TAD boundary prediction tools.**

| Epigenomic data-based methods | | | | |
|---|---|---|---|---|
| cdBEST | Enrichment | http://e-portal.ccmb.res.in/e-space/rakeshmishra/cdBEST.html | Web, Perl | [134] |
| HiCfeat | Penalized multiple logistic regression | https://cran.r-project.org/web/packages/HiCfeat/index.html | R | [135] |
| HiCblock | Generalized linear model | https://cran.r-project.org/src/contrib/Archive/HiCblock/ | R | [136] |
| nTDP | Generalization of Conditional Random Field | https://www.cs.cmu.edu/~ckingsf/research/ntdp/ | | [137] |
| TAD-Lactuca | RF, NN | https://github.com/LoopGan/TAD-Lactuca | Python | [138] |
| preciseTAD | RF | https://bioconductor.org/packages/preciseTAD/ | R | [49] |
| DeepMILO | CNN & RNN | https://github.com/tuantrieu/DeepMILO | Python | [139] |
| PredTAD | GBM | https://github.com/jchyr-sbmi/PredTAD/ | R | [140] |
| **DNA sequence-based methods** | | | | |
| TADBoundaryDetector | CNN+LSTM | https://github.com/lincshunter/TADBoundaryDectector | Python | [141] |
| Deep-loop | CNN | https://github.com/linDing-group/Deep-loop | Python | [142] |

| **Epigenomic data-based methods** | | | | |
|---|---|---|---|---|
| CLNN-loop | CNN+LSTM | https://github.com/HaoWuLab-Bioinformatics/CLNN-loop | Python | [143] |
| **Hybrid methods** | | | | |
| pTADS | RF | https://github.com/chrom3DEpi/pTADS | R | [144] |

**TAD boundary prediction**

Topologically Associating Domains (TADs) represent a higher-order level of chromatin interactions [145]. They represent kilobase-to-megabase size regions on the linear genome that are highly self-interacting [6]. TADs have also been reported to constrain enhancer-promoter communication [15,16] and might be related to genome stability [146]. Boundaries of TADs were found to be enriched in architectural factors such as CTCF, RAD21, SMC3, YY1, and ZNF143 [6,8,147,148,49], and boundary strength correlates with their occupancy [149,150]. Furthermore, distinct patterns of histone modifications [6] and other regulatory marks [42] have also been shown to be present at boundaries. These observations strongly suggest that genome annotation data may be used for TAD boundary prediction [137] (Table 3).

The cdBEST tool was among the first to demonstrate the use of TFBS motifs for boundary prediction in 12 Drosophila species [134]. Trained on experimentally validated boundaries, cdBEST scans a genome using a 750bp window in 10bp increments to calculate fold-enrichment values. The well-known transcription factors, such as BEAF and dCTCF, were among the most predictive motifs. cdBEST defines five types of boundaries, and further demonstrates their association with gene expression differences. A subsequent work demonstrated superior performance of cdBEST over k-means clustering and ChromHMM segmentation and developed a RF model for predicting novel boundaries using modENCODE ChIP-seq data and protein binding motifs [151]. Importantly, this later study used RUS and whole-genome cross-validation, shaping the standard framework for genomics machine learning.

Mourad, R. et al. [135] developed the HiCfeat R package that implements multiple logistic regression for TAD boundary prediction. This model was among the first to use the percent of overlap with TFBSs, instead of binary overlaps. The model was trained on TAD boundaries identified by the HiSeg TAD caller (imbalanced data) and evaluated in cross-validation settings using AUROC. The model uses L1 norm LASSO penalization to select the most predictive features. As expected, Human CTCF, Cohesin, ZNF134, and Polycomb group proteins were positive predictors and P300, RXRA, BCL11A and ELK1 were negative predictors. For Drosophila, BEAF and CP190 were the most predictive predictors, followed by enriched, but not strongly predictive, dCTCF. This work demonstrated that predictive modeling using genome annotations works well across organisms.

In a follow-up work, Mourad, R. et al. [136] introduced a generalized linear model (HiCblock) that allows for estimation of blocking effect of architectural proteins. In addition to genome annotations, the model relies on Hi-C data to model the strength of interactions between loci separated by architectural proteins. The model also requires a distance parameter that specifies the distance between interacting loci. It accounts for both DNA and Hi-C-specific biases, such as GC content, mappability, and fragment length. This work, and the previous work, did not account for class imbalance, relying on 10-fold cross-validation and AUROC to select the best model. Applied to Drosophila and Human Hi-C data, the model reveals known and new findings about the blocking effect of architectural proteins at different distance scales. The authors of this work demonstrate how the integration of genome annotation (one-dimensional) and Hi-C (two-dimensional) data can model the insulating effect of TAD boundary proteins.

Hong and Kim [42] have developed a Position-Specific Linear Model (PSLM) for TAD boundary prediction using overlap count with histone marks, TFBSs, and DNAse hypersensitive sites. The model was trained on TAD boundaries identified by the Directionality Index [6] with the balanced set of negative examples and evaluated using 5-fold cross-validation and AUROC and F1-score metrics. The Population greedy search algorithm (PGSA) was used to select the combination of most predictive features. The authors used the model to identify CTCF, SMC3, RAD21, ZNF143, YY1, DNAse, and H3K36me3 as the most predictive features and distinguished between common and cell type-specific TAD boundaries. Although not implemented as a software tool, this work further demonstrates the importance of different types of genome annotations for predicting TAD boundaries.

Ramírez, F. et al. [152] used several standard classification methods (linear, logistic regression, Gradient Boosting Machine (GBM), and RF) to predict TAD boundaries using the binding affinity of TFBS motifs and DNAse hypersensitive sites. Trained on high-resolution data (0.5Kb Drosophila Kc167 cells Hi-C data) with TAD boundaries predicted using a modified TopDom approach, the models reach >70% sensitivity and specificity in 10-fold cross-validation settings. Feature importance ranking further identifies DNAse as the most important predictor, followed by the expected motifs Beaf32, M1BP, and Ctcf. The authors classified TAD boundaries as promoter-associated or non-promoter-associated, and linked gene expression variability with promoter orientation. They demonstrated that machine learning is able to predict TAD boundaries missed among those detected from Hi-C data only, thus improving the resolution of TAD boundary detection. They also provide an interactive resource, Chorogenome, based on HiCExplorer visualization, to explore TAD structure and epigenomic annotations in Drosophila, Mouse, and Human cell lines.

Sefre & Kingsford developed nTDP [137], a non-parametric approach based on a generalization of Conditional Random Field to predict TAD and interior boundaries using histone modification marks. The class imbalance problem was addressed by reweighting positive and negative examples. The model uses cross-validation and normalized variation of information for performance assessment. The authors defined signatures of four and six histone marks that together, but not individually, accurately predict TAD boundaries. The final trained model can be used to predict TAD boundaries across species and cell types.

TAD-Lactuca [138] predicts TAD boundaries using eight histone marks, CTCF, and sequence information (k-mers). Both epigenomic and sequence information were shown to have complementary contributions to prediction accuracy. Two machine learning algorithms, RF and a 4-layer Multi-Layer Perceptron, were tested on a balanced set of positive and negative TAD boundary examples. Using AUROC and AUPRC measures in 5-fold cross-validation settings, the method was shown to outperform HubPredictor and PGSA and are consistent across resolutions.

preciseTAD [49] was the first tool to predict TAD boundaries at base-level resolution. It uses a RF model to learn TAD-genome annotation associations at the resolution of Hi-C data (detected using Arrowhead [8] or Peakachu [125]) and applies its learned associations to predict the probability of each base being a boundary. After evaluating the performance of chromatin states (histone modification marks, DNAse hypersensitive sites, and TFBSs), the authors found that TFBSs (CTCF, RAD21, SMC3, and ZNF143) were the most predictive. They evaluated four methods for measuring the association between a region and a genomic annotation: signal (sum or average), overlap count, overlap percent, and distance to the region and found the latter to be best performing. For cell lines lacking some TFBS data, data imputed with Avocado [153] has been shown to be a good substitute. Among methods for addressing class imbalance, RUS was shown to perform best. Bases with probability of being a boundary equal to 1 are clustered into preciseTAD boundary regions, and boundary points are identified using Partition Around Medoids. Evaluated in cross-chromosome and cross-cell-line settings, preciseTAD demonstrates improvements to CTCF/RAD21/SMC3 signal around the predicted boundary points. Implemented as an R package, it provides coordinates of predicted boundaries for 60 cell lines.

Deep neural network architectures have also been applied to TAD boundary predictions. Rozenwald, M. et al. [154] compared a linear regression, with and without penalization, with a RNN and with a LSTM to predict TAD boundaries on the Drosophila genome using histone and TFBS marks. The RNN + LSTM captures the sequential order of the genome and distance between genomic annotations. It was shown to outperform all types of linear regression using the weighted mean square error as a performance metric.

DeepMILO [139] uses two types of neural networks, CNN and RNN, to learn sequence features of true TAD boundaries and predict the effect of genomic variants on the probability of boundary formation. True TAD boundaries are defined as having both CTCF and Rad21 signals, considering the convergent direction of CTCF motifs. Evaluated by AUROC, the model outperforms word2vec and boosted trees in distinguishing true vs. non-boundaries.

An interesting development was introduced with PredTAD [140], a tool for predicting differential TAD boundaries. The tool employs a GBM trained on a combination of both genomic and epigenomic features. Genomic features include chromosomal number, location, distance to centromere, gene length, and density. Epigenomic features include histone modifications, DNA binding proteins, and DNA accessibility. PredTAD demonstrates high accuracy in predicting gained, lost, and conserved TAD boundaries when comparing breast cancer cell lines MCF10A, MCF7, and T47D. Interestingly, chromosome number emerges as a significant predictive feature alongside well-known factors like CTCF, RAD21, and SMC3. The model's performance remains robust even without chromosome information and this indicates its versatility. Furthermore, PredTAD outperforms other existing prediction methods such as HubPredictor, PGSA, and TAD-Lactuca, suggesting its efficacy in studying the dynamics of chromatin organization. The tool's ability to predict TAD boundary alterations provides insights into gene regulation, signaling pathway activation, and disease states. In particular, this highlights PredTAD's potential for understanding and targeting 3D chromatin remodeling in breast cancer therapy.

Henderson, J. et al. developed TADBoundaryDetector [141], a novel deep learning approach for predicting TAD boundaries in fruit flies using high-resolution Hi-C data. Through extensive experimentation and hyperparameter optimization, a model comprising three convolutional layers followed by a LSTM layer achieved an impressive accuracy of 96%. Validation techniques such as 10-fold cross-validation were employed, with performance metrics including accuracy, MCC, precision, recall, and F1 score utilized to comprehensively assess model performance. Hyperparameter tuning played a crucial role in optimizing model performance, emphasizing the significance of parameter selection in deep learning architectures. The study also identified several sequence motifs enriched in

TAD boundaries, shedding light on the complex biological mechanisms governing chromatin organization. Comparison with traditional, feature-based models highlighted the superiority of deep learning approaches in TAD boundary prediction. Overall, the findings underscore the potential of deep learning in deciphering genomic architecture and provide valuable insights into regulatory elements governing chromatin organization.

Lv, H. et al. developed Deep-loop [142], a sequence-based CNN model for predicting CTCF-mediated chromatin loops. This model used K-tuple Nucleotide Frequency Component (KNFC), Nucleotide Pair Spectrum Encoding (NPSE), Position Conservation and Position Scoring Function (PCPSF) and natural vector. The authors showed that KNFC, NPSE and PCPSF have strong predictive power across different types of chromatin interaction pairs, while PCPSF is the most informative descriptor to discriminate between true chromatin loops and chromatin non-loops. The model outperforms four other common classification algorithms (SVM, KNN, Naïve Bayes, and LSTM) evaluated by AUROC.

CLNN-loop [143] combines CNN and LSTM (CNN layer, max pooling layer, bidirectional LSTM layer, dropout layer, dense layer, and output layer) for CTCF-mediated chromatin loop prediction using 31 sequence-based features designed with Deep-loop (5 are most predictive, including CTCF motif and sequence conservation). The authors utilized five different sequence encoding schemes to encode DNA sequences: Reverse Complement K-mer (RCKmer), Combination Position Scoring Function (CPSF), NPSE, Combination Position-Specific Tri-Nucleotide Propensity based on single-strand (CPSTNPss), and Combination Position-Specific Tri-Nucleotide Propensity based on double-strand (CPSTNPds). The model performs similar to CTCF-MP and Deep-loop (evaluated by AUROC) and performs well across cell types.

A combination of genome annotation and sequence-based data has also been explored for TAD boundary prediction. Wang, Y. et al. developed pTADS [144], a tool for predicting TAD boundaries and their strength (boundary score) using epigenetic and sequence profile-based signals. Sequence-based features include DNA shape which is calculated by the DNAshape program. Minor groove width, propeller twist, roll, and helix twist are transformed into statistical values like mean, RMSE, median, minimum, maximum, standard deviation, and population deviation. In addition to DNA shape, DNA properties and TFBS motif occurrences are also used. Epigenetic features that are used include chromatin states, TFBSs, histones, replication timing, and nucleosome positioning. The model utilizes a RF method followed by variable selection. Then the selected features are used to train the XGBoost-based model. The authors showed that this approach outperformed other baseline predictors (Ada, GBM, SVMLinear, KNN, NNet, NB, RPART) using 10-fold cross-validation and multiple evaluation metrics including AUROC. LASSO was used to prioritize feature importance. Models that combined TF and histone signals were shown to be the best performing, while known CTCF, Insulator, RAD21, H3K20me1 and PML were enriched at TAD boundaries. Notably, DNA shape and motif occurrences were shown to be least predictive. Different models, trained on TADs detected by three different TAD callers, were shown to robustly prioritize the same features and identify similar boundaries. The authors also demonstrated that pTADS, trained on data from a single cell line, performs well in other cell lines.

**A/B compartment prediction**

Initial observations that genome annotations may be associated with 3D chromatin interactions came from the landmark 2009 paper of Lieberman-Aiden, E. et al. The authors demonstrated that chromatin interactions exhibit one of two long-range contact patterns, dubbed the A and B compartments, suggesting the presence of two spatial neighborhoods. These compartments are typically identified with a principal component analysis (PCA) on the observed over expected count correlation matrices, where the positive/negative values of the first eigenvector are associated with A/B compartments, respectively [5]. The A compartment shows enrichment in expressed genes, DNAse hypersensitive sites, H3K36me3 and H3K27me3 histone modifications. The B compartment shows the opposite enrichment patterns [155–157]. A/B compartments partition a genome into roughly similar proportions of active and inactive regions, thus alleviating the class imbalance problem. These compartments are further refined into the five primary Hi-C subcompartments (A1, A2, B1, B2, and B3, plus B4 on chr19) [8], and further into the 10 SPIN states that have distinct associations with genomic annotations [158]. The A/B compartment classification is frequently used to characterize changes in a genome's activity between conditions [159,160].

Although A/B compartments can be directly predicted from Hi-C data, high-resolution cell type-specific Hi-C data is not always available. This prompted methods development for A/B compartment predictions using broadly available genomic annotations. Fortin and Hansen [161] demonstrated that cell type-specific A/B compartments can be reliably detected from methylation and open chromatin data. Using an approach similar to the original A/B detection methodology (eigenvector analysis of the methylation correlation matrix), they showed the first eigenvector generally corresponds well to the first eigenvector obtained from the observed-to-expected normalized genome contact matrix. These observations link methylation and chromatin accessibility with the formation of higher-order chromatin structures and were consistent across different platforms, e.g., when using Illumina 27K and 450K methylation data, DNAse hypersensitive sites, single-cell ATAC-seq, and whole-genome bisulfite sequencing.

Moore, B. L. et al. developed 3DGenome, a RF model to predict A/B compartments from co-localization with 22 TFBSs and 10 histone marks [162]. This model predicts eigenvectors used to segment a genome into A/B compartments. It outperforms regression-based approaches in cross-cell-line evaluation settings, according to the correlation with the Hi-C derived eigenvectors and AUROC. The authors showed that boundaries of the predicted A/B compartments were enriched in CTCF, YY1, H2A.Z, H3K4me2 and other marks and that these enrichments were cell type-specific.

Di Pierro, M. et al. developed MEGABASE, a neural network that uses ChIP-seq genome annotation data (discretized signal of 84 TFBSs and 11 histone marks) to predict A1, A2, B1, B2, B3, B4 subcompartments [163]. These predictions were subsequently used as an input to an energy landscape model for chromatin organization (MiChroM) to predict the 3D structure of individual chromosomes. The model further demonstrated that genome annotations "encode sufficient information to determine the global architecture of chromosomes and that de novo structure prediction for whole genomes may be increasingly possible".

Notably, Hi-C data itself contains sufficient information to predict A1, A2, B1, B2, B3, B4 subcompartments. Xiong & Ma developed SNIPER [164], a deep neural network consisting of a 9-layer denoising autoencoder and a multi-layer perceptron to predict subcompartments. Trained on Hi-C data from the GM12878 cell line, it outperforms the genome annotation-based predictions of a Gaussian HMM and of MEGABASE. SNIPER-derived predictions correlate well with histone mark signal, replication timing, RNA-seq, and TSA-seq which further supports the association of higher-order chromatin structures with genome annotation data.

**Discussion**

Besides the canonical EPIs, chromatin loops, TAD boundaries, and A/B compartment predictions, other higher-order chromatin structures and their association with genome annotation data remain to be explored. These include, but not limited to, differential TAD/loop boundaries, hierarchical TAD boundaries (sub-TADs nested within larger TADs) [165–169], ultra-high resolution 3D genomic features [157], and multi-way chromatin interactions [126,130].

The importance of differential TAD/loop boundaries lies in the fact that 3D genome organization varies between cell-types [5,8], along different stages of the cell-cycle [170–172] even within homogenous populations of synchronized cells [172]. Approximately 60-80% of TADs remain invariant across cell types (constitutive) [173], while the others correspond to flexible structures (facultative) [174,6]. Differential TAD/loop boundaries can be detected by a simple overlap analysis or in a statistically principled way [175–179]. In addition, other differential structures, such as differential A/B compartments [159] and differential stripes [180], can be better defined. Despite the biological relevance and interpretability of such 3D changes, methods for predicting them, and consequently the knowledge of associated genomic features, remain undeveloped and a highly promising direction for future investigation.

Different types of TAD/loop boundaries are similarly not considered by prediction methods. Not all TAD boundaries are created equal [181] and TAD boundary strength correlates with occupancy of architectural proteins [149,150]. TADs can form hierarchies, with sub-TADs nested within larger TADs [165,166,8,167,168]. With the advent of ultra-high-resolution technologies such as Micro-C (nucleosome-resolution) [182], we anticipate the development of methods to predict the strength and hierarchy of TAD/loop boundaries.

Prediction of multi-way chromatin interactions has been hindered by limited technology and data availability. Several recently developed technologies started to close this gap. Genome Architecture Mapping (GAM) [183] quantifies chromatin contacts by sequencing DNA from a set of ultrathin nuclear sections at random orientations. Trac-looping [184] captures multiscale contacts by inserting a transposon linker between interacting regions. DNA SPRITE [185] follows a split-pool procedure to assign unique barcodes to individual complexes, with read pairs sharing identical barcodes capturing multi-way chromatin interactions. As the multi-way chromatin interaction data and standards for its representation become more developed, we anticipate the development of methods predicting multi-way interactions.

The utility of DNA sequence features for chromatin interaction predictions remains to be explored [186]. Besides one-hot encoding and k-mer representations of DNA sequence, recent methods introduce additional methods for feature extraction. The iLearnPlus [187] web-based tool includes 147 unique feature sets capturing various properties of DNA/RNA/protein sequences as well as 21 machine-learning algorithms with 7 deep-learning approaches for their extraction, clustering, normalization, and predictor construction. It remains to be seen how informative such features actually are for predicting TAD/loop boundaries or other chromatin interaction patterns.

Another promising avenue for improving predictions of 3D organization is by using gene expression data. It has been shown that distal methylation, open chromatin regions, TFBSs, and histone modifications can predict gene expression [188–190]. Conversely, incorporating gene expression in chromatin network analysis can improve active enhancer prediction [191].

The pervasive problem with many chromatin interaction prediction methods remains the lack of consensus on "ground truth" or "gold standard" for reference chromatin interaction data. Some studies use databases such as FANTOM, while others rely on external tools that define chromatin interactions from Hi-C maps. A more biologically justifiable approach is to use experimentally obtained data such as proximity-ligation ChIA-PET [192], PLAC-Seq [193], HiChIP [194], Capture-C [195], and/or Capture Hi-C [196]. Recently developed high-throughput imaging approaches such as STORM [197] and HiFISH [198] can directly measure spatial distances at the single-cell level. To increase the robustness and interpretability of chromatin interaction prediction methods, viable methods must define ground truth chromatin interactions with imaging approaches.

## Summary

In this review, we summarize the use of machine learning methods to predict various 3D features of genomes. These methods draw on data from different sources such as DNA sequence features, genome annotation data, histone modifications, and DNAse hypersensitivity sites. We cover methods for predicting promoter-enhancer interactions, chromatin looping interactions, TAD boundaries, and A/B compartments. We discuss the challenges associated with this type of analysis including class imbalance and the need for careful validation. We also highlight the importance of using appropriate metrics to assess performance. Overall, we provide a comprehensive overview of the current state-of-the-art machine learning tools to gain insights into the 3D organization and function of genomes.

*Conflict of Interest.* None.

## Funding

This work was supported in part by the George and Lavinia Blick Research Scholarship to MD and the NIH/NCI (R01CA246182, R21CA273779) grants to JCH.